\documentclass[aps,prc,twocolumn,groupedaddress,showpacs]{revtex4}

\usepackage{graphicx}

\begin{document}
\title{Similarity between nuclear rainbow  and  meteorological rainbow\\
 - evidence for  nuclear  ripples -
  }

\author{
S. Ohkubo$^{1}$   and   
Y. Hirabayashi$^2$}
\affiliation{$^1$ Research Center for Nuclear Physics, Osaka University, 
Ibaraki, Osaka 567-0047, Japan }
\affiliation{$^2$Information Initiative Center,
Hokkaido University, Sapporo 060-0811, Japan}

\date{\today}

\begin{abstract}
We present   evidence for  the nuclear  ripples superimposed on the 
 Airy structure of  the nuclear rainbow, which is   similar to  the meteorological rainbow.  
  The mechanism of the nuclear  ripples is  also similar
 to that of the meteorological rainbow, which is caused by the interference
 between the externally  reflective waves  and refractive waves.  The nuclear  ripple structure  was
 confirmed by analyzing the elastic angular distribution in  $^{16}$O+$^{12}$C rainbow 
 scattering at 
$E_L$=115.9 MeV  using the  coupled channels method by  taking  account of  coupling
 to the excited states of  $^{12}$C and  $^{16}$O with   a double
 folding model  derived from a density-dependent effective nucleon-nucleon force  with 
 realistic wave  functions for $^{12}$C and $^{16}$O. 
 The coupling to the  excited states
plays the role of creating  the external reflection.
  \end{abstract}

\maketitle

Descartes \cite{Descartes} and subsequently Newton  \cite{Newton} explained the  rainbow  
in optics by reflection and refraction in the raindrops.   Airy \cite{Airy} understood  the
 supernumerary rainbow by the   wave nature of light. 
The mechanism of the meteorological rainbow was understood precisely only recently by Nussenzveig
 using the electromagnetic theory of light  \cite{Nussenzveig1992}.
In analogy with the meteorological rainbow the nuclear rainbow was predicted theoretically
\cite{Ford1959} and observed in $\alpha$ particle scattering   \cite{Goldberg}.
The rainbow has been observed  also  in other systems such as in atom-atom
collisions, atom-molecule  collisions \cite{Connor1981},  electron-molecule collisions 
\cite{Ziegler1987} and atom scattering from crystal surfaces \cite{Kleyn1991}.
Although the  mechanism of  Newton's zero-order ($p=1$ in Fig.~1)   nuclear rainbow \cite{Michel2002}, 
 where only  refraction is  active is very different
 from that of the meteorological rainbow ($p=2$ in Fig.~1), a similar Airy structure has been 
observed.  As shown in Fig.~2, the precise description  of the meteorological
 rainbow  given  by solving Mie scattering shows the rapidly oscillating  structure, 
 the high-frequency   ripple structure,  superimposed on the Airy structure of the rainbow
 \cite{Beeck1996,Lee1998,Nussenzveig1992}. 
The ripple  structure is not predicted  by the semiclassical theory of the nuclear rainbow of Ref. \cite{Ford1959}
and no attention has been paid to its  possible existence.
 Here we report for the first time  evidence for the  existence  of the ripple structure in the
 observed nuclear rainbow and explain its mechanism.
 
The Airy structure of nuclear rainbows has been studied extensively
  especially for  heavy ion scattering such as    $^{16}$O+$^{16}$O,
 $^{16}$O+$^{12}$C  and $^{12}$C+$^{12}$C  \cite{Khoa2007,Brandan1997}.
For the most  typical   $^{16}$O+$^{16}$O system, 
similar to the typical  $\alpha$+$^{16}$O and $\alpha$+$^{40}$Ca   
  scattering \cite{Michel1998,Ohkubo1999},  a  global deep potential has been 
 determined uniquely  from the rainbow scattering. It
  reproduces the experimental data over a wide
 range of energies  from negative energy to  the incident energy $E_L$=1120 MeV -  that is,  
 the  rainbows  \cite{Khoa2000}, prerainbows \cite{Nicoli1999}, molecular resonances and 
 cluster   structures with the superdeformed configuration \cite{Ohkubo2002} -
 in a unified way.  Unfortunately  the observed 
Airy structure in the angular distributions is obscured due to symmetrization
of two identical bosons. 

   In this respect  rainbow scattering  of the asymmetric $^{16}$O+$^{12}$C  system
is important  and has been thoroughly investigated 
\cite{Brandan1986,Khoa1994,Ogloblin1998,Nicoli2000,Ogloblin2000,Szilner2001}.
A global deep potential could describe well the   rainbows   in 
the high-energy region,   prerainbows \cite{Michel2002}, molecular resonances, and 
 cluster structures with the  $^{16}$O+$^{12}$C  configuration 
in the quasibound energy region  in a unified way \cite{Ohkubo2004}. 
However at  energies around  $E_L$=100 MeV 
  the global  optical  potential   calculations \cite{Nicoli2000}    only  reproduced   the
 experimental  angular distributions  in a   qualitative way  at larger angles.
Also to reproduce the high-frequency  oscillations   imaginary potentials  - with 
 a  thin-skinned   volume  term  and    an extraordinary  small diffuseness parameter  around 
0.1 fm   accompanying a surface term peaked at a  larger radius - were  needed 
\cite{Nicoli2000,Szilner2001}. 
 
The purpose of this paper is to show that the high-frequency  oscillations
 superimposed  on the Airy structure  are nothing but the ripple structure of the nuclear
 rainbow  and can be
 explained by  fully taking account of  coupling to the excited  states of $^{12}$C and $^{16}$O
  by using the microscopic  wave functions  and the  extended  double folding model. 
The mechanism of the ripple structure  and the role of coupling to the excited states 
  is clarified,  and the similarity between the macroscopic meteorological rainbow and the 
 quantum nuclear rainbow, despite  the difference of the underlying interactions,  is
 discussed.  
 \par

We study    $^{16}$O+$^{12}$C  scattering  with the coupled channels method  using 
an extended  double folding (EDF)  model that describes all the diagonal and off-diagonal
coupling potentials derived from  the microscopic   realistic wave functions for $^{12}$C  
and $^{16}$O  using  a density-dependent   nucleon-nucleon force.
  The diagonal and coupling potentials for the $^{16}$O+$^{12}$C system are calculated using
 the EDF  model  without introducing a normalization factor:
\begin{eqnarray}
\lefteqn{V_{ij,kl}({\bf R}) =
\int \rho_{ij}^{\rm (^{16}O)} ({\bf r}_{1})\;
     \rho_{kl}^{\rm (^{12}C)} ({\bf r}_{2})} \nonumber\\
&& \times v_{\it NN} (E,\rho,{\bf r}_{1} + {\bf R} - {\bf r}_{2})\;
{\it d}{\bf r}_{1} {\it d}{\bf r}_{2} ,
\end{eqnarray}
\noindent where $\rho_{ij}^{\rm (^{16}O)} ({\bf r})$ is the diagonal ($i=j$)  or transition ($i\neq j$)
 nucleon  density of  $^{16}$O    taken from  the microscopic $\alpha$+$^{12}$C  cluster model 
 wave functions calculated  in  the orthogonality 
condition model (OCM) in Ref.\cite{Okabe1995}. The model uses  a  realistic size parameter both 
 for the $\alpha$ particle and $^{12}$C,  and is an extended version of  the  
OCM $\alpha$ cluster model  of Ref. \cite{Suzuki1976}, which  reproduces almost  all the energy levels  
well up  to $E_x$$\approx$13 MeV and the  electric transition probabilities  in $^{16}$O. 
We take into account  the important transition densities 
available in Ref.\cite{Okabe1995}, i.e., g.s. $\leftrightarrow$  $3^-$ (6.13 MeV) and 
$2^+$ (6.92 MeV)  in addition to all the
 diagonal  potentials.  
$\rho_{kl}^{\rm (^{12}C)} ({\bf r})$ represents the diagonal ($k=l$) or transition ($k\neq l$)
 nucleon density of $^{12}$C  calculated using the microscopic three-$\alpha$ cluster model 
in the resonating group method \cite{Kamimura1981}. This model reproduces the structure of 
 $^{12}$C  well,  and the  wave functions     have  been checked for many experimental  data 
\cite{Kamimura1981}. In the coupled channels calculations we  take into account  
the   0$^+_1$ (0.0 MeV), $2^+$ (4.44 MeV),   and 3$^-$ (9.64 MeV) states of $^{12}$C.
 The mutual excitation channels in which both   $^{12}$C and $^{16}$O are excited simultaneously
 are not   included.  For the  effective interaction   $v_{\rm NN}$     we use  
 the DDM3Y-FR interaction \cite{Kobos1982}, which takes into account the
finite-range nucleon  exchange effect.
 An imaginary potential (nondeformed) is introduced   phenomenologically to take into account the effect
of absorption due to other channels.

\begin{figure}[t]
\includegraphics[keepaspectratio,width=8.0cm] {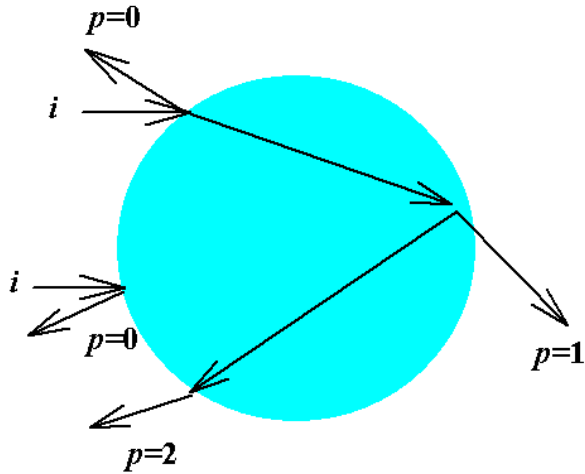}
\caption{ Illustrative figure of path of an  incident ray ($i$) in a spherical raindrop 
in geometrical optics. The rays  $p=0$,  $p=1$ and $p=2$ correspond to reflection, 
refraction only, and the primary rainbow, respectively.  }
\label{fig1}
\end{figure} 

\begin{figure}[t]
\includegraphics[keepaspectratio,width=8.0cm] {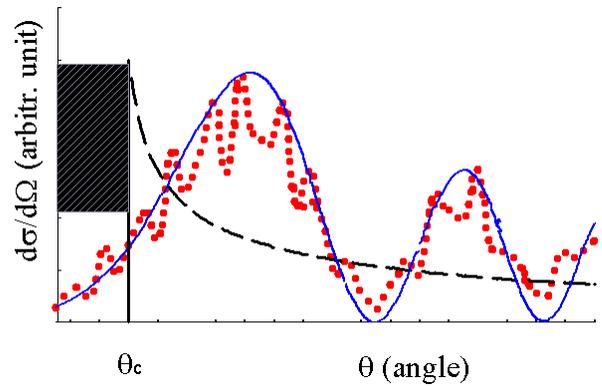}
\caption{Cross sections for the primary rainbow in Fig.~1. The dashed line is by classical theory 
in optics and the dark region before the critical angle ($\theta_c$=138$^\circ$) is displayed by 
a black shade.
The dotted and solid   lines show calculations of Mie scattering  \cite{Lee1998} and in the Airy 
approximation  \cite{Lee1998}  in the wave theory of light, respectively.  }
\end{figure}

In Fig.~3(a) the angular distributions of elastic $^{16}$O+$^{12}$C  scattering at 
$E_L$=115.9 MeV   calculated  using the coupled channels method (blue solid line)
  are compared with
 the experimental data.
  We found that  the  EDF potential works well without introducing a normalization factor.
  The volume integral per nucleon pair  of the ground state diagonal part, 
 $J_V$=317.7 MeVfm$^3$, is consistent with those used in other   optical potential model
  calculations  and     belongs to the same global potential family  found in the 
$E_L=$62$-$1503 MeV region  \cite{Nicoli2000,Szilner2002,Ogloblin2000}.
 The  parameters used in  the imaginary potential with a  Woods-Saxon volume-type form factor
displayed in Fig.~4  are    $W_V$= 14 MeV,   $R_W$= 5.6 fm, and   $a_W$= 0.20  fm with 
a volume integral  per nucleon pair   $J_W$=54.3 MeVfm$^3$,  in the conventional notation. 
 We see that  the refractive farside scattering dominates at the intermediate and large angles.
The calculation reproduces well  the  two broad  Airy maxima in the angular range, 
$\theta$=60$-$90$^\circ$ (Airy maximum $A2$) and
 $\theta$=100$-$140$^\circ$  (Airy maximum $A1$) 
   in the experimental angular distribution, which are brought about by the  refracted farside 
component.  
 Also the high-frequency oscillations superimposed on the two broad Airy maxima, 
$A1$ and $A2$, 
  in the experimental data  are  reproduced well.  We see that the
 high-frequency oscillations    are brought  about by the interference between the  farside  and 
 nearside scattering components.  The  investigation of the contributions of  each channel
 reveals that  none is overwhelmingly dominant and that  the contribution of  the excited states
 of $^{16}$O  is  as much  as that  of   $^{12}$C, which is quite different from the higher energy
 region  around $E_L$=300 MeV where  coupling to the $2^+$ state of $^{12}$C  contributed
 dominantly  in creating the secondary bow in the classically forbidden darkside of the primary 
rainbow \cite{Ohkubo2014}.
 Also neither the extremely thin-skinned    volume-type imaginary potential with $a_W$=0.1 
 nor the  surface 
 imaginary potential  peaked at a larger   radius used in Refs. \cite{Nicoli2000,Szilner2001} were
  needed.

\begin{figure}  [thbp]
\includegraphics[keepaspectratio,width=8.6cm] {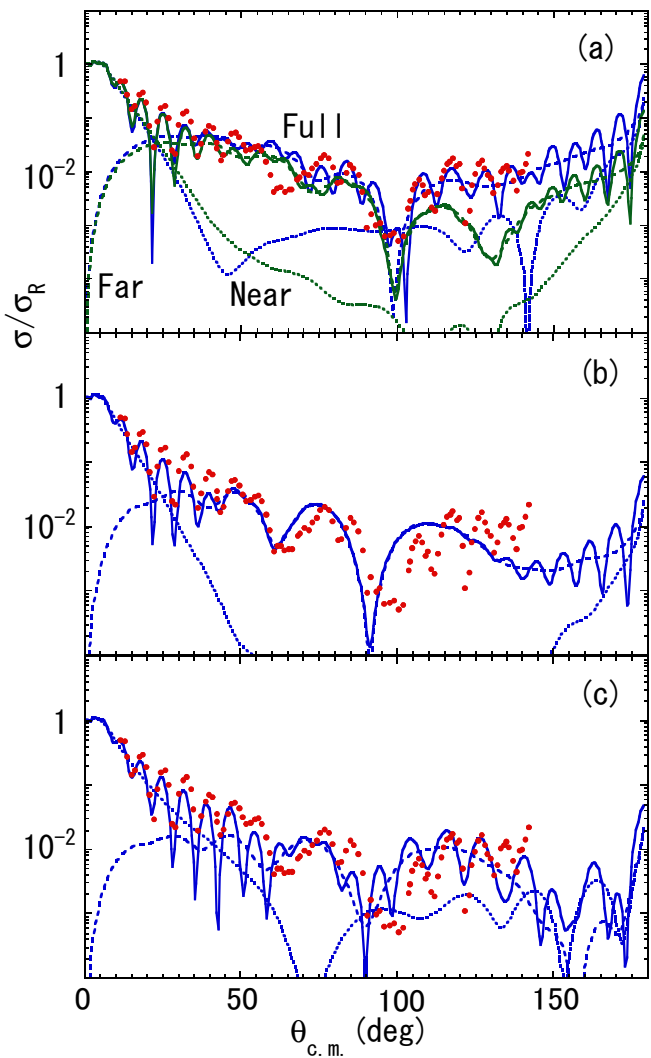}
 \protect\caption{\label{fig.3} {(Color online)   
The  experimental cross sections (points) in  $^{16}$O+$^{12}$C  scattering at $E_L$=115.9 MeV 
 \cite{Nicoli2000} are compared with the calculations  (solid line) using  the EDF potential:
 (a)  the coupled channels calculations  with $a_W$=0.2 (blue line), 
   (b) the single channel calculations with $a_W$=0.6, 
and  (c) the single channel calculations using the  extremely  thin-skinned 
 volume-type   imaginary potential with $a_W$=0.1. 
For comparison, in (a) the coupled channels calculations with  $a_W$=0.4  are displayed by
the green line. The calculated cross sections (solid line) are
 decomposed into the farside  (dashed line) and  nearside (dotted line) components.
 }
}
\end{figure}

In Fig.~3(b) the  angular distributions calculated  in the single 
 channel calculation using the readjusted imaginary potential (displayed in Fig.~4),
 $W_V$=11.5 MeV,  $R_W=$5.9 fm and  $a_W=$0.6  fm ($J_W$=52.8 MeVfm$^3$),
 which is similar to  Ref. \cite{Szilner2002}, 
  are shown.  Although the farside scattering is dominant, similar
 to Fig.~3(a), and  the gross behavior of the Airy structure of the experimental 
 angular distribution is reproduced, the high-frequency oscillations
  are missing.  By comparing   Fig.~3(a) (blue solid line)  and Fig.~3(b), we note
 that the channel coupling to the excited states of $^{16}$O and $^{12}$C 
 contributes in generating
the high-frequency oscillations, although the rather  small $a_W$=0.2 is  needed.
  In  Fig.~3(b) we note that the  nearside component is damped  more
 than two order of magnitude compared with the farside  component, and  no
 interference between them occurs resulting no high-frequency oscillations. 
  While the  $J_W$ values are almost  the same for Fig.~3(a) and 3(b),  the nearside scattering
is retained  significantly  in Fig.~3(a). This means that the   channel coupling in  Fig.~3(a) 
 plays a role of increasing  the nearside scattering  component, i.e., reflection.

\begin{figure}  [t]
\includegraphics[keepaspectratio,width=5.0cm] {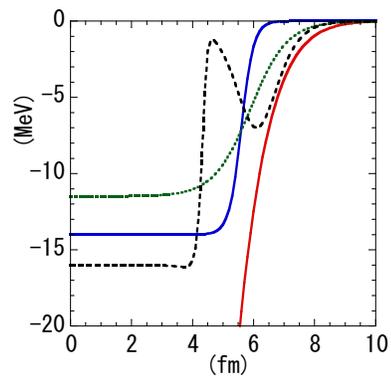}
 \protect\caption{\label{fig.4} {(Color online)   The real folding  and Woods-Saxon
 volume-type imaginary potentials 
used  in the coupled channels calculations [Fig.~3(a)] in $^{16}$O+$^{12}$C  scattering 
 are displayed by the red solid line and the blue solid line, respectively.
The  imaginary potentials used in the single calculations 
in Fig.~3(b) and 3(c)  are displayed by the green dotted  line and the black dashed line, respectively.
}
}
\end{figure}

\begin{figure}  [b]
\includegraphics[keepaspectratio,width=6cm] {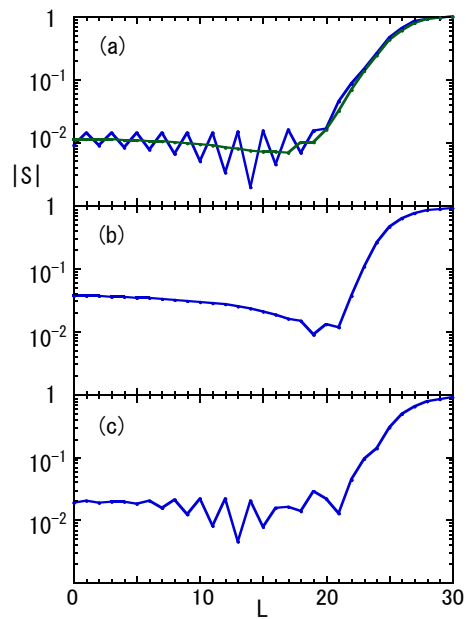}
 \protect\caption{\label{fig.3} {(Color online)  
 The moduli of the calculated $S$ matrices in Fig.~3(a), 3(b) and 3(c).  
The lines are explained  in the caption of  Fig.~3. }
}
\end{figure}

 This can be confirmed in  Fig.~3(c)  where the angular distributions calculated in the single
 channel  using both the  extremely thin-skinned volume-type (Woods-Saxon squared) 
imaginary potential  with  $W_V=$16 MeV, $R_W$=4.4 fm, and $a_W=$0.1 fm
  and the surface imaginary     potential with  $W_D$=7 MeV, $R_D$=6.1 fm, and 
$a_D$=0.46 fm ($J_W$=59.8 MeVfm$^3$)  (displayed in Fig.~4),  which are  similar to those 
in  Refs. \cite{Nicoli2000}   and \cite{Szilner2001},   are shown.
We note that the values of  $J_W$ are almost the same for the three 
cases (a), (b), and (c).
 The   high-frequency oscillations superimposed on the Airy structure are recovered only by using
 this extremely thin-skinned imaginary potential.
We see that the nearside component needed to  bring about the high-frequency oscillations  
is significantly increased compared with Fig.~3(b).
The  sharper  the diffuseness    of   the imaginary potential is, the more   the   nearside
 component is increased. This can be checked   by decreasing 
 the ratio  $a_W$/$W_V$; that  is,   if we increase the strength of the imaginary potential
 in Fig.~3(c) by 50 $\%$ to $W_V$=24 MeV   from 
the original 16 MeV, the magnitude of the calculated   cross section  and its  farside component 
decrease  as expected from the increase of  absorption. 
However,  the magnitude of the  nearside  component  is {\it  increased}. 
This means that for the nearside scattering  the imaginary potential  does not act
 as absorption but acts as ``divergence,''  i.e., increasing reflective waves under the small 
diffuseness $a_W$=0.1.  Thus the increased nearside component is found to be reflective in  origin.
The same behavior is observed  for  the coupled channels calculations in  Fig.~3(a) with
 $a_W$=0.2.   We see in Fig.~3(a) that the calculated cross sections with a moderate
smooth diffuseness  $a_W$=0.4 (green line) show  no high-frequency oscillations, which
 means that no reflective waves are  created.
 We note that the 
 magnitudes of the $S$ matrix  in Figs.~5(a) (blue line)  and 5(a) (green line)  are 
 similar to those in  Figs.~5(c)  and    5(b), respectively.

\begin{figure}  [htb]
\includegraphics[keepaspectratio,width=8.6cm] {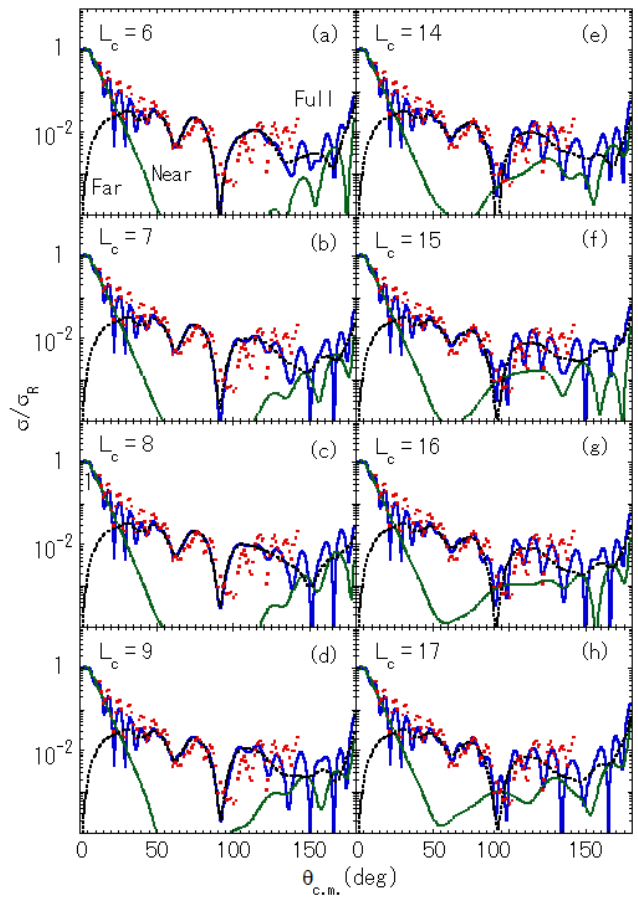}
 \protect\caption{\label{fig.6} {(Color online)   
The angular distributions in  $^{16}$O+$^{12}$C  scattering at $E_L$=115.9 MeV 
 calculated by replacing the $S_L$ with $L=0\sim L_c$ 
 among the $S_L$-matrices generated  by the potential  used for Fig.~3(b)   by those
 generated by the potential used for Fig.~3(c) (solid line) and its  farside  (dashed line) 
and   nearside (dotted line) components are compared with the experimental data (points)
  \cite{Nicoli2000}. 
}
}
\end{figure}

We show that the high-frequency oscillations superimposed on the Airy structure in Fig.~3(a) are
nothing but the ripple structure of the nuclear rainbow.
In Fig.~2 the ripple structure in the meteorological rainbow is  generated 
by the interference between  the $p=0$  external  direct reflection 
and the  $p=2$  refractive rainbow rays with one internal reflection \cite{Nussenzveig1992}.
However  it has been considered that  in  the nuclear rainbow, which is caused by a  Luneberg lens \cite{Michel2002} 
of  a  nuclear  potential  ($p=1$)  with a smoothly diffused  surface,   direct reflection scarcely
 occurs.  This is seen  in  Fig.~3(b)  where  no high-frequency oscillations appear  in the calculations
 using  the real  and  imaginary potentials with    a smooth surface.
On the other hand, we see in Fig.~3(c) that  the high-frequency oscillations   are created by 
the interference between the reflective nearside  component  caused by the   sharp-edged
imaginary potential and the farside  component.
 Thus the high-frequency oscillations superimposed on the Airy structure are considered  to be 
the  nuclear ripples  because they have the same physical origin as those of 
 the meteorological rainbow.
In Fig.~3(a) 
it is found   that  the coupling to the excited states  also contributes in  creating
 the reflective nearside  waves that are caused
 by the sharp-edged   imaginary potential  in Fig.~3(c).  
The nearside scattering waves that are responsible for the generation of the high-frequency
 oscillations in Figs.~3(a) and 3(c)   correspond to the externally reflected waves  in Mie scattering 
of the meteorological rainbow.
Now  the physical meaning and the origin of the extremely thin-skinned volume-type imaginary 
potential needed  in Refs. \cite{Nicoli1999,Szilner2001} are clear. They were needed to  mimic
 the effect of the  channel coupling  to enhance the    reflective  waves.
 
 In the single channel calculations, the increase of the nearside component that corresponds to the
externally reflective waves is only attained by
 using  the  extraordinary small  diffuseness parameter $a_W$=0.1 for the imaginary potential. 
This    necessarily  accompanies  introducing   the additional surface imaginary  potential at the 
large radius to  preserve the net absorption, i.e., the volume integral of the imaginary potential.
In the coupled channel calculations in Fig.~3(a),  in fact,  no surface  imaginary potential at the large
 radius was needed.
It is   important  to treat channel    coupling to the
excited states microscopically   
to avoid the unphysically sharp-edged volume-type imaginary potential   and the 
  surface  imaginary potential at the large radius.

How the ripple structure emerges on the Airy  structure by the reflective waves is shown
 in Fig.~6 where  the angular distributions -  calculated by replacing the $S_L$ matrices with 
 the orbital angular momentum  $L=0\sim L_c$ 
  generated  by the potential  of Fig.~3(b)   by those 
 generated by the potential of  Fig.~3(c)  -  are displayed.
With around $L_c$=8 (corresponding impact parameter $b=$2.1  fm) the  ripple structure starts 
to emerge on the $A1$ Airy peak in the backward 
angles beyond $\theta\approx 100^\circ$   and with  around $L_c$=14  ($b=$3.6  fm) 
they also appear on the Airy
 peak $A2$  in the intermediate angles before $\theta\approx 90^\circ$.
  For $L_c=14-17$ ($b=$3.6-4.3  fm), which corresponds to the radius of the sharp
 edge  of the imaginary potential in Fig.~4 (black dashed line),  the high-frequency oscillations in 
phase  with the experimental data are reproduced. 
The impact parameters of these partial waves are significantly  smaller than  
$b=$5.9  fm of the grazing partial wave $L=24$  for which $|S_L|$=0.5 (see Fig.~5).  
The channel coupling in Fig.~3(a) plays  two roles; that is,   enhancing the reflective waves 
 and enhancing   absorption at the surface,  which makes it
 possible  to use the relatively larger diffuseness parameter $a_W$=0.2 fm  
and no surface imaginary potential  needed in Fig.~3(c).

 Finally we mention that the high-frequency oscillations are not due to the
 elastic transfer  of the $\alpha$ particle  \cite{Szilner2002}.
 In fact, we see  in the 
  detailed  coupled  reaction channels calculations of  $^{16}$O+$^{12}$C  scattering at 
$E_L$=115.9 MeV in Ref.  \cite{Rudchik2010}  that
 the contribution of the elastic transfer is three orders of magnitude
 smaller than the experimental data.
  Also the  present  calculations  take into account the one-nucleon exchange effect, which is suggested 
to prevail over other transfer reactions \cite{Rudchik2010},  by using the effective  interaction 
DDM3Y, in which   the knock-on exchange   effect is incorporated \cite{Kobos1982,Brandan1997}.

To summarize, we have calculated  $^{16}$O+$^{12}$C scattering with the 
 Airy structure  at $E_L$=115.9 MeV using a coupled channels
 method with an extended  double folding (EDF) potential that is derived by using  the microscopic realistic wave functions 
for $^{12}$C and $^{16}$O by  taking  account of  excited states of   
the  $2^+$ (4.44 MeV)   and 3$^-$ (9.64 MeV) states of $^{12}$C and the  3$^-$ (6.13 MeV) and  
  2$^+$ (6.92 MeV) states of $^{16}$O.
 Our calculations reproduce the high-frequency oscillations
 superimposed on the Airy structure.
 It is found that  the high-frequency oscillations are nothing but the nuclear ripples similar to those
 superimposed on the
 Airy structure in Mie scattering of the   meteorological  rainbow. The nuclear ripples are generated 
by the interference between the  refractive
 waves and the externally  reflected  waves.    The coupling to the excited states of 
$^{16}$O and $^{12}$C plays the role of creating  external reflection. 
 Although the  active  interactions in the nuclear 
 and the meteorological rainbows are very different, we see the similarity 
 in that  both have the ripple structure on the Airy structure due to the
 same origin of the interference between refractive waves and the externally reflected waves.
It is startling that a classical concept of  a ripple in the meteorological rainbow persists
in the quantum nuclear rainbow. 

One of the authors (S.O.) thanks the Yukawa Institute for Theoretical Physics for
 the hospitality extended  during a stay in Spring 2014. 
Part of this work was  supported by the Grant-in-Aid for the Global COE Program ``The Next
 Generation of Physics, Spun from Universality and Emergence'' from the Ministry 
of Education, Culture, Sports, Science and Technology (MEXT) of Japan.


\begin{thebibliography}{aa}
\bibitem {Descartes}
R.  Descartes, 
Le Discours de la methode (sous-titre Pour bien conduire sa raison, et chercher la verit\'{e}
 dans les sciences) plus la Dioptrique, Les Meteores et la Geometrie (Leiden,  1637).
\bibitem {Newton}
I.  Newton, {\it Opticks or, a Treatise of the Reflexions, Refractions, Inflexions and 
Colours of Light.}  (London, 1704); (Dover, New York, 1952).
\bibitem{Airy}
 G.  B. Airy,  {Trans. Camb. Philos. Soc.} {\bf 6},  379 (1838).
\bibitem {Nussenzveig1992}
H. M.  Nussenzveig,
Sci. Am. {\bf 236}, 116 (1977);
H. M. Nussenzveig,
{\it Diffraction Effects in Semiclassical Scattering}
(Cambridge University Press, Cambridge, 1992). 
\bibitem{Ford1959}
K. W. Ford and  J. A.  Wheeler, Ann  Phys. {\bf 7},  259 (1959).
\bibitem {Goldberg}
D. A. Goldberg and  S. M.  Smith, 
 Phys. Rev. Lett. {\bf 29}, 500 (1972);
D. A. Goldberg,
 S. M.  Smith,  H. G. Pugh,  P. G.  Roos, and  N. S. Wall,  
 Phys. Rev. {\bf C7}, 1938 (1973);
 D. A. Goldberg,
, S. M. Smith,  and G. F.   Burdzik, 
 Phys. Rev. C  {\bf 10}, 1362 (1974).


\bibitem {Connor1981}
J. N. L. Connor and D. Farrelly, 
J. Chem. Phys.  {\bf  75},  2831 (1981).
\bibitem {Ziegler1987}
G. Ziegler,
 M. R\"{a}dle, O. P\"{u}tz, K. Jung, H. Ehrhardt, and K. Bergmann,
Phys. Rev. Lett. {\bf 58}, 2642 (1987).

\bibitem {Kleyn1991}
A. W. Kleyn and T. C. M. Horn, 
Phys. Rep. {\bf 199}, 191 (1991);
C. O. Reinhold,
 J. Burgd\"{o}rfer, K. Kimura, and M. Mannami, 
Phys. Rev. Lett. {\bf 73}, 2508 (1994).
\bibitem {Michel2002}
F. Michel, G. Reidemeister, and  S.   Ohkubo, 
Phys. Rev. Lett. {\bf 89}, 152701 (2002).

\bibitem {Lee1998}
R. L. Lee, Jr.,
Appl.  Opt. {\bf  37},   1506 (1998). 
\bibitem {Beeck1996}
J. P. A. J. van Beeck and M. L. Riethmuller,
Appl.  Opt. {\bf  35},   2259 (1996). 


\bibitem {Khoa2007} 
D. T. Khoa,  W. von Oertzen, H. G. Bohlen, and S.  Ohkubo, 
J. Phys. {\bf G 34}, R111 (2007), and references therein.
\bibitem {Brandan1997}
 M. E.  Brandan and   G. R.  Satchler,
Phys. Rep. {\bf 285}, 143  (1997). 
\bibitem {Michel1998}
F. Michel,  S.  Ohkubo, and G.  Reidemeister, 
   Prog. Theor. Phys. Suppl. {\bf 132}, 7 (1998). 
\bibitem {Ohkubo1999}  
S. Ohkubo,  T.  Yamaya, and P. E.   Hodgson,
 Nuclear clusters. in {\it Nucleon-Hadron Many-Body
 Systems}, 
(edited by  H.  Ejiri and  H. Toki (Oxford University Press, Oxford, 1999),  p. 150  .


\bibitem {Khoa2000}
D. T. Khoa, W. von Oertzen, H. G. Bohlen, and F. Nuoffer,
 Nucl. Phys.  {\bf A672},   387 (2000).
\bibitem {Nicoli1999}
M. P. Nicoli  {\it et al.},
 Phys. Rev. C {\bf 60}, 064608 (1999).
\bibitem {Ohkubo2002} 
 S. Ohkubo and K.  Yamashita, 
 Phys. Rev. C {\bf 66}, 021301(R) (2002);
S. Ohkubo, Heavy Ion Physics (Acta Physica Hungarica A) {\bf18}, 287 (2003).

 \bibitem {Brandan1986}
M. E. Brandan  {\it et al.},
 Phys. Rev. C {\bf 34}, 1484  (1986).
\bibitem{Khoa1994} 
D. T. Khoa, W. von Oertzen,  and  H. G.   Bohlen, 
 Phys. Rev. C  {\bf  49}, 1652  (1994).

\bibitem {Ogloblin1998}
A. A. Ogloblin  {\it et al.}, 
Phys. Rev. C {\bf 57}, 1797 (1998).

\bibitem {Nicoli2000}
M. P. Nicoli  {\it et al.}, 
Phys. Rev. C  {\bf  61},  034609 (2000).
\bibitem {Ogloblin2000}
A. A. Ogloblin  {\it et al.}, 
 Phys. Rev. C {\bf 62}, 044601 (2000).
\bibitem {Szilner2001}
S. Szilner  {\it et al.}, 
 Phys. Rev. C {\bf  64}, 064614  (2001).


\bibitem {Ohkubo2004} 
S. Ohkubo and  K.  Yamashita,  
 Phys.  Lett. {\bf B578}, 304 (2004).


\bibitem {Okabe1995} 
S.  Okabe, 
{\it Tours Symposium on Nuclear Physics II}, edited
by H. Utsunomiya {\it et al.} (World Scientific, Singapore, 1995),
p. 112.
\bibitem {Suzuki1976} 
Y.~Suzuki,
Prog. Theor. Phys. {\bf 55}, 1751 (1976);
 Prog. Theor. Phys. {\bf 56}, 111 (1976). 
\bibitem{Kamimura1981}
 M. Kamimura,
Nucl. Phys. {\bf A351}, 456 (1981).

\bibitem {Kobos1982}
A. M. Kobos,
 B. A. Brown, P. E. Hodgson,  G. R. Satchler, and A. Budzanowski,
 Nucl. Phys. {\bf A384}, 65 (1982);
A. M. Kobos,
 B. A. Brown, R. Lindsaym, and G. R. Satchler,
Nucl. Phys. {\bf A425}, 205 (1984).
\bibitem {Szilner2002}
S. Szilner, 
 W. von Oertzen, Z. Basrak, F. Haas, and M. Milin,
Eur. Phys. J. {\bf A 13}, 273 (2002).
\bibitem {Ohkubo2014}
 S.   Ohkubo and Y. Hirabayashi, 
Phys. Rev.  C {\bf 89}, 051601(R) (2014).


\bibitem {Rudchik2010}
A. T. Rudchik {\it et al.}, 
Eur. Phys. J. {\bf A  44}, 221 (2010).

\end{thebibliography}
\end{document}